\documentclass[onecolumn,showpacs,amsmath,amssymb,prl]{revtex4}
\usepackage{epsfig}
\usepackage[a4paper]{geometry}
\usepackage[in]{fullpage}
\newcommand{\beq}{\begin{equation}}
\newcommand{\eeq}{\end{equation}}
\newcommand{\bea}{\begin{eqnarray}}
\newcommand{\eea}{\end{eqnarray}}
\begin{document}
\title{ Testing Hadronic Models of Gamma Ray Production at the Core
 of Cen A} 
\author
{Jagdish C. Joshi \footnote{jagdish@rri.res.in} and Nayantara Gupta  \footnote{nayan@rri.res.in} }
\affiliation
{Astronomy \& Astrophysics Group, Raman Research Institute,\\ Bangalore 560 080, India}
\begin{abstract}

Pierre Auger experiment has observed a few cosmic ray events above 55 EeV from the direction of the core of Cen A. These cosmic rays might have originated from the core of Cen A. High energy gamma ray emission has been observed by HESS from the radio core and inner kpc jets of Cen A. 
We are testing whether pure hadronic interactions of protons or heavy nuclei with the matter in the core region or photo-disintegration of heavy nuclei can explain the cosmic ray and high energy gamma ray observations from the core of Cen A. The scenario of $p-\gamma$ interactions followed by photo-pion decay 
has been tested earlier by Sahu et al. (2012) and found to be consistent with the observational results. In this paper we have considered some other possibilities (i) the primary cosmic rays at the core of Cen A are protons and the high energy gamma rays are produced in $p-p$ interactions,(ii) the primary cosmic rays are Fe nuclei and the high energy gamma rays are produced in $Fe-p$ interactions and (iii) the primary cosmic rays are Fe nuclei and they are photo-disintegrated at the core. The daughter nuclei de-excite and high energy gamma rays are produced.
The high energy gamma ray fluxes expected in each of these cases are compared with the flux observed by HESS experiment to normalise the spectrum of the primary cosmic rays at the core. We have calculated the expected number of cosmic ray nucleon events between 55 EeV and 150 EeV in each of these cases to verify 
the consistencies of the different scenarios with the observations by Pierre Auger experiment. We find that if the primary cosmic rays are Fe nuclei then their photo-disintegration followed by de-excitation of daughter nuclei may explain the observed high energy particle emissions from the core of Cen A. The luminosity of the cosmic ray Fe nuclei required to explain the observational results of HESS and Pierre Auger is higher than the luminosity of the cosmic ray protons in the $p\gamma$ interaction model. The required cosmic ray luminosity depends on the density of the low energy photons at the source which photo-disintegrate the Fe nuclei and the size of the emitting region.

\end{abstract}
\date{\today}
\maketitle
\section{Introduction}
Directional correlation of ultrahigh energy cosmic rays (UHECRs) and their possible sources has long been studied with the events from Volcano Ranch \cite{vol}, SUGAR \cite{sugar}, Fly's Eye \cite{fly}, HiRes \cite{hires}, AGASA \cite{agasa}, Yakutsk \cite{yak}, Haverah Park \cite{hav}, AUGER \cite{auger1} experiments. 
Compact radio quasars were correlated with UHECRs in \cite{fb}. UHECR events have been correlated earlier with their sources assuming small angular deflections \cite{dub,tiny1,tiny2}. Galactic magnetic field may play a crucial role in revealing the charge composition of UHECRs \cite{tiny3}. BL Lacertae objects have been correlated with UHECR events by different authors. While Gorbunov et al. \cite{gor} suggested $\gamma$-ray emission by BL Lacertae objects could be related to their capability of UHECR emission, the study by Torres et al. \cite{tor} showed the correlation of UHECRs with BL Lacertae objects is not significant. Correlation of clustered events from AGASA experiment with BL Lacertae objects and decay of massive relic particles \cite{blasi} was examined \cite{evans} and negative result was reported. The correlation of compact radio quasars or 3EG gamma-ray blazars and the cosmic ray events above 10EeV was studied with the available data at that time in \cite{sigl} and the authors concluded there is no significant correlation.
Virmani et al. \cite{vir} found angular correlation of cosmic ray events above 100 EeV with radio-loud compact QSO sources. 
\par
The results of correlation studies are highly dependent on the samples of data used from different experiments. This field has remained exciting as different groups came up with different conclusions. The more interesting aspect is there may be exotic physical phenomena which may lead to the detection of particles from sources beyond the distance limit due to Greisen-Zatsepin-Kuzmin (GZK) \cite{gzk} attenuation. For example violation of Lorentz invariance \cite{col,gupta} and axion like particles or exotic massive hadrons \cite{alb} may lead to the detection UHECRs from far away sources. With the successful operation of Pierre Auger experiment many UHECR events have been detected which could be successful to shed light on the correlation of UHECR events above 55 EeV with nearby sources \cite{auger1,auger2}.
Pierre Auger experiment has measured the anisotropy of 69 UHECR events above energy 55 EeV and correlated with the positions of AGN within $3.1^{o}$ and distance 75 Mpc from us. The degree of the observed correlation has decreased from earlier study \cite{auger2,auger1}.
\par 
More analysis with future data may establish whether UHECR sources are associated with luminous infrared galaxies \cite{ber1}. In a recent paper \cite{zaw} it has been discussed that the correlation of the highest energy UHECR events from the Pierre Auger expriment with AGN (Active Galactic Nuclei) in V\'eron-Cetty-V\'eron catalogue is stronger than the correlation with the random sets of galaxies selected from large-scale structure. This result implies that the correlation of UHECRs with AGN is not entirely due to AGN tracing the distribution of matter in large-scale structure.
Chandra observations on some of the AGN correlated with UHECR events from Pierre Auger experiment were used to confirm whether their X-ray and optical emissions support strong nuclear activities \cite{ter}. Ten galaxies were studied and none of them showed significant AGN component.
This study reduces the number of correlated AGN with UHECR events observed by Pierre Auger experiment. However more data are needed to confirm the results of this study. Another study on the correlation of local AGN with UHECR events observed by Pierre Auger experiments further reveals that the claimed correlation should be considered as a result of chance coincidence \cite{mos}. 
UHECRs may come from steady/continuous sources or flares or transients. Advances in radio astronomy have opened the opportunity of observing a large number of transients and exploring the dynamic sky. Radio observations have revealed that mildly relativistic supernovae may accelerate cosmic rays to energies above 60 EeV \cite{sayan}.
\par

The region around our closest radio-galaxy Cen A shows the largest concentration of events relative to the isotropic expectations. Cen A being our nearest radio-galaxy at a distance of 3.4 Mpc has been studied extensively as a potential source of UHECRs \cite{anch1,anch2,gopal1,bier1}. More investigations in this direction would be helpful to understand the sources of UHECRs.
\par
 Its TeV gamma ray emission \cite{hess} has been related to the two extremely energetic cosmic ray events observed by Pierre Auger from the direction of the core of Cen A \cite{sahu} within the hadronic model of $p\gamma$ interactions. In this scenario the luminosity of the cosmic ray protons has to be close to the Eddington luminosity of the black hole, which is $L_{Edd}\sim 1.3\times 10^{46} (M/10^{8}M_{sun})erg/sec$. In another paper it has been discussed that \cite{anch2} a cosmic ray luminosity of $9\times 10^{39}$ erg/sec is required in the energy interval of 55 EeV and 150 EeV to explain these two events observed by Pierre Auger experiment in 3 years. If the lower energy bound is 1 EeV then the estimated luminosity in the band of 1 EeV to 150 EeV is $5\times10^{40}$erg/sec, which is below the bolometric luminosity $10^{43}$ erg/sec of this source.
\par
Fraija et al. \cite{fraija} have explained the emission of gamma rays observed by HESS and the UHECR events observed by Pierre Auger with $p-p$ interactions assuming the gamma ray emission is from the lobes of Cen A. They have shown that the scenario of $p-\gamma$ interactions in their model is not consistent with the observational results.   
\par
 The composition of the primary cosmic rays inside their sources is not known, this allows us to assume different compositions. The observed diffuse UHECRs may be a mixture of protons and heavy nuclei \cite{auger3}. With increasing energy the difficulty of determining the composition of the observed diffuse UHECRs increases.
\par
The composition of the cosmic ray events observed from the direction of Cen A is not known. They could be protons or neutrons or heavy nuclei. 
We have assumed that the two cosmic ray events observed from the direction of the core of Cen A are protons or neutrons and they are emitted from the core of Cen A as it has been assumed in \cite{sahu}.

High energy gamma rays can be produced in interactions of the primary cosmic ray protons with cold matter protons at the core ($p-p$), in $p-\gamma$ interactions followed by decay of photo-pions, pure hadronic interactions of cosmic ray heavy nuclei with cold matter protons (Fe-p) and photo-disintegration of cosmic ray heavy nuclei followed by de-excitation of daughter nuclei. 
In the paper by Sahu et al. \cite{sahu} they have assumed the primary cosmic rays are only protons and the gamma rays are produced in $p-\gamma$ interactions.
They have shown that the scenario they have considered is consistent with
the observational results.
In this work we consider some other possibilities. 
We have studied the following scenarios or possiblities  
 (i) the primary cosmic rays are only protons and the high energy gamma rays are produced in $p-p$ interactions,
(ii) the primary cosmic rays are only Fe nuclei and their interactions with the cold matter protons ($Fe-p$) lead to the production of the high energy gamma rays and (iii) the primary cosmic rays are Fe nuclei, they are photo-disintegrated
and the high energy gamma rays produced from the de-excitation of the daughter nuclei. 
We have compared the calculated gamma ray flux in each of these cases with the one observed by HESS experiment.
Thus observed high energy gamma ray flux from Cen A is useful to reveal the hadronic processes inside this source \cite{nayan1,kachel}. 
Finally, we check the consistency of each of these scenarios with the observations by Pierre Auger experiment.
\section{UHECRs and Gamma Rays from Cen A}
 Fermi collaboration has observed the gamma ray emission from the core of Cen A \cite{abdo}. The higher energy gamma rays observed from Cen A by HESS 
\cite{hess} could be more useful to study this source as a UHECR accelerator.  
HESS experiment has observed gamma ray emission from the radio core and the inner kpc jets \cite{hess}. The gamma ray flux above energy 250 GeV is a single power law with index $2.73\pm 0.45_{stat}\pm0.2_{sys}$,
  denoted as 
$\frac{d\phi_{\gamma}^o(E_{\gamma}^o)}{dE_{\gamma}^o dt^o dA}$ in observer's frame on earth. 
\beq
\frac{d\phi_{\gamma}^o(E_{\gamma}^o)}{dE_{\gamma}^odt^odA}=2.45\times 10^{-13}\Big(\frac{E_{\gamma}^o}{1TeV}\Big)^{-2.73} cm^{-2} sec^{-1} TeV^{-1}
\label{hess}
\eeq
The very high energy gamma ray emission observed by HESS is from the core and inner jets of Cen A.

\subsection{Pure Hadronic Interactions}
  In pure hadronic interactions ($p-p$, $A-p$ where $A$ is the mass number of the heavy nucleus) neutral and charged pions $\pi^0$, $\pi^+$ and $\pi^-$
are produced with almost equal probabilities. Neutral pions decay to high energy gamma rays and the charged pions to neutrinos and antineutrinos. 
In the first case as mentioned earlier we have assumed that the primary UHECRs at the core of Cen A are only protons. Their optical depth for pion production in interactions with hydrogen of molecular density $n_H cm^{-3}$ in a blob of size of $R=3\times10^{15}$cm in the wind rest frame \cite{abdo} is $\tau_{pp}=R/l_{pp}$, where the mean free path is $l_{pp}= 3/n_H\times 10^{25}cm$, for interaction cross-section $\sigma_0=34.6$mb.Each pion produced in $p-p$ interactions is assumed to carry $20\%$ of the initial proton's energy.
 
We have calculated the high energy gamma ray flux in the observer's frame on earth to compare with HESS observations.  

 The Dopler factor $\delta_D$ is $\delta_D=\Gamma^{-1}(1-\beta cos{\theta_{ob}})^{-1}$.
$\beta$ is the dimensionless speed of the wind rest frame with respect to the observer on earth and the angle between the observed photon and the wind's velocity is $\theta_{ob}$ as measured in observer's frame. $\Gamma$ is the Lorentz boost factor of the wind rest frame. Cosmic rays can be emitted in any direction in the wind rest frame as well as the gamma rays produced in their interactions. 
There will be beaming in the observer's frame. The photons travelling along our line of sight are observable. The photons detected on earth has a Doppler shifted energy which depends on their angle of emission with respect to the direction of the velocity of the wind rest frame and $\Gamma$.
The deflection of ultrahigh energy cosmic ray protons with energy more than 56 EeV is on the average $3^{o}$ in Galactic magnetic field. Cen A is 3.4 Mpc away from us. The deflection of the cosmic ray protons of energy more than 56 EeV is negligible in extragalactic magnetic field in this case \cite{auger1}.   
 The cosmic ray proton/neutron events detected above 55 EeV energy with directionality within $3^{o}$ of the core of Cen A are travelling from the source to the observer with the same Doppler shift in energy as the gamma rays observed by HESS if they all have a common origin.  
 The energies and times in the observer's frame and wind rest frame are related as $E_{\gamma}^o=\delta_D E_{\gamma}$ and $t^{ob}=t/\delta_D$, we have neglected the redshift correction as redshift ($z$) of Cen A is much less than 1.
The gamma ray flux expected from decaying energetic pions produced in interactions of UHECR protons (expressed in number of protons per unit energy per unit time $\frac{dN_p}{dE_p dt}(E_p)$ in wind rest frame) with matter \cite{anch3,nayan2} at the core region of Cen A is
\beq
\frac{d\phi_{\gamma}^o(E_{\gamma}^o)}{dE_{\gamma}^o dt^o dA}=\frac{2Y_{\alpha}}{\bf{4\pi D^2}}\frac{R}{l_{pp}}\int_{E_{\pi^0,min}}^{E_{\pi^0,max}}\frac{dN_p(E_{\pi^0})}{dE_{\pi^0}dt } \frac{dE_{\pi^0}}{(E_{\pi^0}^2-m_{\pi^0}^2)^{1/2}}.
\label{pp_gamma}
\eeq
In the above equation the number of cosmic ray protons per unit energy at the core of Cen A $\frac{dN_p(E_{\pi^0})}{dE_{\pi^0}dt }=A_pE_{\pi^0}^{-\alpha}$, $A_p$ is the normalisation constant and $\alpha$ is the spectral index. The distance to the source is $D=3.4Mpc$.
 The minimum energy of pions is $E_{\pi^0,min}=E_{\gamma}+m_{\pi^0}^2/(4E_{\gamma})$ and the maximum energy is $E_{\pi^0,max}=0.2 E_{n}^{max}$ where $E_{n}^{max}$ is the maximum energy of cosmic ray proton/nucleon,$m_{\pi^0}$ is pion's rest mass and $E_{\gamma}$ is the energy of gamma rays.   
 The spectrum weighted moment $Y_{\alpha}$ has been calculated from \cite{anch3}.
\beq
Y_{\alpha}=\int_0^1 x^{\alpha-2} f_{\pi^0}(x)dx
\label{y_al}
\eeq
The function $f_{\pi^0}(x)\simeq 8.18 x^{1/2}\Big(\frac{1-x^{1/2}}{1+1.33x^{1/2}(1-x^{1/2})}\Big)^4\Big(\frac{1}{1-x^{1/2}}+\frac{1.33(1-2x^{1/2})}{1+1.33x^{1/2}(1-x^{1/2})}\Big)$.
For $\alpha=2.73$ we get $Y_{\alpha}= 0.03$. With eqn.(\ref{hess}), eqn.(\ref{pp_gamma}) and eqn.(\ref{y_al}) we can find the normalisation constant of the UHECR proton spectrum $A_p$. UHECR neutrons produced in $p-p$ interactions subsequently decay to protons, electrons and antineutrinos. We have also included the UHECR neutrons decaying to protons in calculating the expected UHECR event rate in Pierre Auger. The integrated exposure of the Pierre Auger detector is $(9000/\pi) km^2$ and relative exposure for declination angle ($\delta=47^{o}$) is $\omega(\delta)\simeq 0.64$. The number of UHECR events expected in Pierre Auger detector can be calculated using the UHECR spectrum.
 The UHECR spectra in observer's frame and wind rest frame are related as
\beq
\frac{dN^o_{p,n}(E^o_{p,n})}{dE^o_{p,n}dt^o dA}=\frac{1}{4\pi D^2}
\frac{dN_{p,n}(E_{p,n})}{dE_{p,n}dt}
\eeq
and the number of expected events is
\beq
N_{p,n}^o= \frac{15}{4} \times \frac{9000}{\pi}(km^2) \omega(\delta)\int_{E_{l}^o}^{E_{u}^o}  \frac{dN_{p,n}^o(E_{p,n}^o)}{dE_{p,n}^o dt^o dA }dE_{p,n}^o
\eeq
We have used $E^o_{p,n}=\delta_D E_{p,n}$ as we have calculated the expected number of events in Pierre Auger which travelled in the direction $\theta_{ob}$.
Also, we have assumed $\delta_D=1$ which corresponds to $\Gamma=7$ and $\theta_{ob}=30^o$.   
In the above equation the lower and upper limits of the energy bin are $E_l^o=55$EeV and $E_u^o=150$EeV respectively.
If we assume that the proton spectral index remains 2.73 upto the highest energy and they are not deflected by the intervening magnetic field then in 15/4 years 450 events are expected for $\tau_{pp}=10^{-6}$, which corresponds to $n_H=10^4 cm^{-3}$. For lower densities $\tau_{pp}$ will be smaller. In this case many more protons may escape from the source before interacting with the matter near the core region. The intervening magnetic field may deflect them away from us and some of them travelling towards us would trigger the detectors at the Pierre Auger observatory. As we are predicting a very large number of UHECR events in this case, the scenario of $p-p$ interactions at the core is not favoured by the observational data from Pierre Auger. 
In the $p-p$ interaction scenario the luminosity of UHECRs in the energy bin of 55EeV and 150EeV is estimated as $L_{UHECR}\simeq3\times 10^{43}/n_H$ erg/sec, which is much less than the Eddington luminosity $L_{Edd}=10^{46}$ erg/sec for Centaurus A. 

\par
 In the second case we have assumed that the primary cosmic rays are only Fe nuclei and they are interacting with cold matter protons at the core region of Cen A. In this case 
 
 the rate for $Fe-p$ interactions is $R_{Fe-p}= n_H \sigma_{Fe} c $, where the cross-section for interaction of nuclei of mass number 56 is $\sigma_{Fe}=34.6\times 56^{3/4}mb$.
If UHECRs are Fe nuclei then pure hadron interactions may lead to the production of gamma rays. The cross-section of interactions are $A^{3/4}$ times higher in comparision to p-p interaction and hence the rate of $A-p$ interactions is also higher by the same factor.
 If we consider there are only iron nuclei near the core region of Cen A, then the gamma ray flux expected on earth in pure hadron interactions $Fe-p$ is
\beq
\frac{d\phi_{\gamma}^o(E_{\gamma}^o)}{dE_{\gamma}^o dt^o dA}=\frac{2Y_{\alpha}}{{\bf 4\pi D^2 }}\frac{R}{l_{Fep}}\int_{E_{\pi^0,min}}^{E_{\pi^0,max}}\frac{dN_{Fe}(E_{\pi^0})}{dE_{\pi^0}dt } \frac{dE_{\pi^0}}{(E_{\pi^0}^2-m_{\pi^0}^2)^{1/2}}.
\label{Fep_gamma}
\eeq
 The number of UHECR Fe nuclei per nucleon energy per unit time at the core region of Cen A is,
$\frac{dN_{Fe}(E_p)}{dE_{p}dt}=56\frac{dN_{Fe}(E_{Fe})}{dE_{Fe}dt}$, with $E_{Fe}=56E_p$. We have expressed the number of Fe nuclei per unit energy of neutral pions per unit time as $\frac{dN_{Fe}(E_{\pi^0})}{dE_{\pi^0}dt}$.
The mean free path of $Fe-p$ interactions has been denoted by $l_{Fep}$, where 
$l_{Fep}=0.048 l_{pp}$.
Eqn.(\ref{Fep_gamma}) can be expressed as 
\beq
 \frac{d\phi_{\gamma}^o(E_{\gamma}^o)}{dE_{\gamma}^odt^o dA}=\frac{2Y_{\alpha}}{\bf{ 4\pi D^2}}56^{-\alpha+1}\frac{R}{l_{Fep}}\int_{E_{\pi^0,min}}^{E_{\pi^0,max}}\frac{dN_{p}(E_{\pi^0})}{dE_{\pi^0}dt} \frac{dE_{\pi^0}}{(E_{\pi^0}^2-m_{\pi^0}^2)^{1/2}}.
\eeq
In pure hadron interactions protons or neutrons will be produced with 
 neutral or charged pions respectively. We calculate the flux of nucleons (protons and neutrons) produced in pure hadron interactions.
\beq
E_{p,n} \frac{dN_{p,n}(E_{p,n}) }{dE_{p,n}dt} dE_{p,n}=
0.8\frac{R}{l_{Fep}}E_{Fe}\frac{dN_{Fe}(E_{Fe})}{dE_{Fe}dt} dE_{Fe}
\eeq
where, $E_{p,n}= 0.8E_{Fe}/56$, assuming the secondary nucleon takes away $80\%$ of the primary nucleon's energy.
In this case the secondary nucleon flux produced in $A-p$ interactions is very low and we expect no event in Pierre Auger detector in $15/4$ years.
 Hence, we conclude that neither $p-p$ nor $Fe-p$ interaction scenario is consistent with the observational results from the core of Cen A. 
 \subsection{Photo-Disintegration of Heavy Nuclei}
Photo-disintegration process of gamma ray emission has been discussed in many earlier papers \cite{steck,anch3}. 
 If the primary cosmic rays are only Fe nuclei at the core of Cen A then
 they may be photo-disintegrated by the low energy photons in that region. 
The multi-wavelength observations have revealed the broad band spectral energy distribution (SED) at the core of Cen A as shown in our FIG. 2 \cite{abdo}.
After the photo-disintegration of the primary nuclei daughter nuclei and secondary nucleons (protons/neutrons) are produced. The daughter nuclei de-excite by emitting gamma rays. If the observed high energy gamma ray emission from Cen A is due to this process then we can calculate the expected nucleon (proton/neutron) flux from Cen A using the observed gamma ray flux \cite{anch3}.
 The rate of photo-disintegration process is calculated with eqn.(6) of \cite{anch3}
\beq
R_{phot-dis}=\frac{c\pi \sigma_0 \epsilon_0'\Delta}{4\gamma_p^2} \int_{\epsilon_0'/2\gamma_p} ^\infty  \frac{dn(x)}{dx}\frac{dx}{x^2}.
\eeq
The value of the cross-section normalization constant is $\sigma_0=1.45Amb$, the central value of GDR $\epsilon_0'=42.65 A^{-0.21}MeV$ for $A>4$ and width of the GDR is $\Delta=8 MeV$. 
The Lorentz factor of each nucleon is $\gamma_p=E_{Fe}/(56 m_p)$.
We have used the photon spectral energy distribution observed on earth 
${\epsilon_{\gamma}^o}^2\frac{dN_{\gamma}^o(\epsilon_{\gamma}^o)}{d\epsilon_{\gamma}^odt^odA}(MeV cm^{-2} sec^{-1})$ from the fit given in \cite{abdo} also shown with solid curve in our FIG.2.
The photon density per unit energy in the core region $\frac{dn(x)}{dx}$ is
\beq
4\pi R^2 c \frac{dn(x)}{dx}=4\pi D^2\delta_D^{-p}\frac{dN_{\gamma}^o(\epsilon_{\gamma}^o) }{d\epsilon_{\gamma}^o dt^o dA}\eeq
where, $p=n+\alpha+2$ and $n=2,3$ for continuous and discrete jet respectively 
\cite{gis}. We have denoted the energy of the low energy photons in the observer's frame by $\epsilon_{\gamma}^o$ and $\epsilon_{\gamma}^o=\delta_D x$.
$\alpha$ is the spectral index of the SED given in FIG.2., ${\epsilon_{\gamma}^o}^2\frac{dN_{\gamma}^o(\epsilon_{\gamma}^o)}{d\epsilon_{\gamma}^odt^odA}
\propto {\epsilon_{\gamma}^o}^{-\alpha}$. 
$\alpha$ takes different values in different energy regimes as shown in FIG.2.
From the above equation it is noted that the photon density at the source depends on $\delta_D$. In Abdo et al. \cite{abdo} they have taken various values of $\Gamma$ and $\delta_D$, the SED fit of SSC model to Fermi data is given for $\Gamma=7$ and $\delta_D=1$ which corresponds to $\theta_{ob}=30^{o}$. 
For smaller values of $\delta_D$ the photon density at the source would be much higher.
The distance of the source $D=3.4 Mpc$ and the radius of the core region $R=3\times 10^{15}cm$.
In photo-disintegration process protons and neutrons can be produced with equal probabilities. TeV gamma rays may be produced in this process from PeV UHECRs.
Similar to eqn.(28) given in \cite{anch3} we can relate the neutron, proton and gamma ray fluxes from photo-disintegration of nuclei of mass $A$. 
\beq
\frac{dN_{n,p}^o(E_{n,p}^o)}{dE_{n,p}^odt^o dA}=\frac{\bar E'_{\gamma A}}{ m_n \bar n_A} \frac{d\phi_{\gamma}^o(E_{\gamma}^o)}{dE_{\gamma}^odt^o dA}\label{phot_dis_nucleon}
\eeq
where in the wind rest frame $E_{\gamma}=E_n\bar E'_{\gamma A}/m_n$. 
We are interested to calculate the number of proton or neutron events in Pierre Auger above 55 EeV which maintain their directionality while travelling from the core of Cen A to the observer. 
They have the same Doppler shift in energy as the gamma rays observed by HESS as they are produced in the same wind frame and travelling in the same direction from the source to the observer.
If the gamma ray emission is monochromatic in the rest frame of the nucleus then its average has been denoted by $\bar E'_{\gamma A}$. $\bar n_A$ is the average multiplicity of gamma rays and $m_n$ is rest mass of each nucleon.
For Fe nuclei $ \bar E'_{\gamma 56}=2-4$ MeV and gamma ray multiplicity is $\bar n_{56}=1-3$. Assuming the same spectral index of the neutron and proton spectrum from TeV to the highest energy we calculate the expected number of events in Pierre Auger detector in 15/4 years in the energy bin of 55EeV to 150EeV.
We get two events for spectral index 2.45 with $\bar E'_{\gamma,56}=4$ MeV and $\bar n_{56}=2$ which agrees with the detection by Pierre Auger experiment from the direction of the core of Cen A. The power law spectrum which fits HESS data has spectral index $2.73\pm0.45_{stat}\pm0.2_{sys}$ \cite{hess}. The spectral index $2.45$ used in our calculations is within the range of error in the spectral index obtained by HESS group.

In this scenario variability of the source increasing the emission may yield
more UHECR events from the direction of Cen A. Due to the low gamma ray flux from Cen A it was not possible by HESS experiment to detect variabilities in time scales shorter than days and with increments below a factor of 15-20 \cite{hess}.  
If the size of the emission region is $R=3\times10^{15}$ cm \cite{abdo}, and the rate of the photo-disintegration process is $R_{phot,dis}$ then the high energy gamma ray emission can be related to the number UHECR Fe nuclei per nucleon energy per unit time at the core of Cen A $\frac{dN_{Fe}}{dE_N dt}(E_N)$ as follows
\beq
\frac{d\phi_{\gamma}^o(E_{\gamma}^o)}{dE_{\gamma}^o dt^o dA}=\frac{1}{\bf{4\pi D^2}}\frac{R}{\beta c}\frac{\bar n_{56} m_N}{2 \bar E'_{\gamma,56}}\int_{\frac{m_N  E_{\gamma}}{\bar 2E'_{\gamma,56}}} \frac{dN_{Fe}(E_N)}{dE_N dt} R_{phot,dis} \frac{dE_N}{E_N}
\label{phot_dis_iron}
\eeq
where $\beta=v/c \sim 1$ for UHECR nuclei. 
  
We calculate the normalization constant of the UHECR Fe nuclei flux from eqn.(\ref{phot_dis_iron}). 
The HESS spectrum is measured above $E_{\gamma}^o=250 GeV$. Gamma rays of energy 250 GeV are produced by Fe nuclei of per nucleon energy $E_N=E_{\gamma} m_N/(2\bar E'_{\gamma,56})=29$ TeV.

In our FIG.1. we have plotted the rate of photo-disintegration of Fe nuclei with the energy per nucleon in the wind rest frame along x-axis. Between 1 TeV to 100 TeV nucleon energy in the wind rest frame the rate is almost constant and it is $2\times10^{-8} sec^{-1}$, for $\Gamma=7$ and $\delta_D=1$. At higher energy the rate increases but the cosmic ray nuclei flux decreases more rapidly as it follows a power law with spectral index $-2.45$.

In this case the luminosity of the UHECR Fe nuclei flux in the energy bin of $55\times 56$EeV and $150\times 56$EeV is $\sim10^{42} erg/sec$ which is much below the Eddington's luminosity.
The 170 KeV photons at the second peak of SED in FIG.2. photo-disintegrate Fe nuclei of energy $E_{Fe}=2.8 TeV$. This result is obtained using the threshold energy condition $\epsilon'_0/2\gamma_p=170 KeV$, where $\gamma_p$ is the Lorentz factor of each nucleon in wind rest frame and we have used $\delta_D=1$ for the Doppler shift of the low energy photons. 
The gamma ray energy produced from photo-disintegration of 2.8 TeV Fe nuclei is
calculated using the expression $ E_{\gamma}=2 \bar E'_{\gamma,56} E_N/m_N$,
 where $\bar E'_{\gamma,56}=4 MeV$ and energy of each nucleon $E_N=50 GeV$.
We find the peak energy in the gamma ray spectrum from photo-disintegration of Fe nuclei by $170 KeV$ photons is at $400 MeV$.
The spectrum of cosmic ray Fe nuclei has a break at 2.8TeV due to the second 
peak in the SED at 170 KeV. Above 2.8 TeV the spectral index -2.45 gives a 
good fit to the observational results. The total luminosity of the Fe cosmic rays has to be of the order of $10^{47}erg/sec$, which is higher than the Eddington's luminosity of Cen A.
We note that the luminosity required to accelerate cosmic rays to above $10^{20}$eV in Cen A is higher than $10^{46} erg/sec$ \cite{dermer,abdo}. Dermer et al. \cite{dermer} have shown that the apparent isotropic luminosity can easily exceed $10^{46} erg/sec$ in Cen A during high flaring states for small beaming cones. 
\par
The SED we have used is the fit to SSC model obtained by Abdo et al. \cite{abdo}. There are error bars on the observed photon flux and also there are no observational data points between the two peaks as shown in our FIG.1. The lower energy photons photo-disintegrate the higher energy Fe nuclei. The rate of photo-disintegration is directly proportional to the density of low energy photons at the source. Higher density of low energy photons would lead to higher rate of photo-disintegration process. If the rate of photo-disintegration is higher then a lower luminosity of cosmic rays would be required to explain the observational 
results.

\par
Along the direction of the x-ray jet of Cen A the x-ray photon density is higher. 
This would lead to more efficient production of high energy gamma rays and require lower UHECR luminosity.

\par
We have shown the high energy gamma ray spectrum from photo-disintegration of Fe nuclei in FIG.2 with a black dashed line. 
\section{$p\gamma$ Interactions}
The $p\gamma$ interaction rate is calculated using the photon SED from 
 \cite{abdo} and the formalism dicussed in \cite{wax}.
\beq
R_{p\gamma}=\frac{c}{\gamma_p^2}\int_{\epsilon_o}^{\infty} \sigma(\epsilon)
\xi \epsilon d\epsilon \int_{\epsilon/(2\gamma_p)}^{\infty} \frac{dn(x)}{dx}\frac{dx}{x^2}
\eeq
where $\sigma(\epsilon_{peak})=0.5mb$ is the cross section of interaction at the resonance energy $\epsilon_{peak}=0.3GeV$ in the proton rest frame and the full width of the resonance at half maxima is $0.2 GeV$. The fractional energy going to a pion from a proton is $\xi=0.2$. The threshold energy of pion production in proton rest frame is $\epsilon_o=0.15 GeV$. $p\gamma$ process has been discussed in detail in \cite{sahu}. They have shown that it can explain the observational results. In this model the luminosity of the cosmic ray protons at 13 TeV has to be $4\times10^{45}$erg/sec for production of 190 GeV gamma rays. The optical depth for $p\gamma$ interactions for 13 TeV protons with 170 KeV photons is estimated to be $10^{-6}$ in \cite{sahu}. We get similar optical depth for $p\gamma $ interactions at 13 TeV proton energy using our calculated rate of $p\gamma$ interactions given in FIG.1. 
\section{Discussions and Conclusions}

Our calculated rates of the various processes of high energy gamma ray production are shown in FIG.1. with hydrogen density $n_H=1.7 cm^{-3}$ and photon spectral energy distribution (SED) from \cite{abdo}. 
\begin{figure*}[t]
\centerline{\includegraphics[width=3.25in]{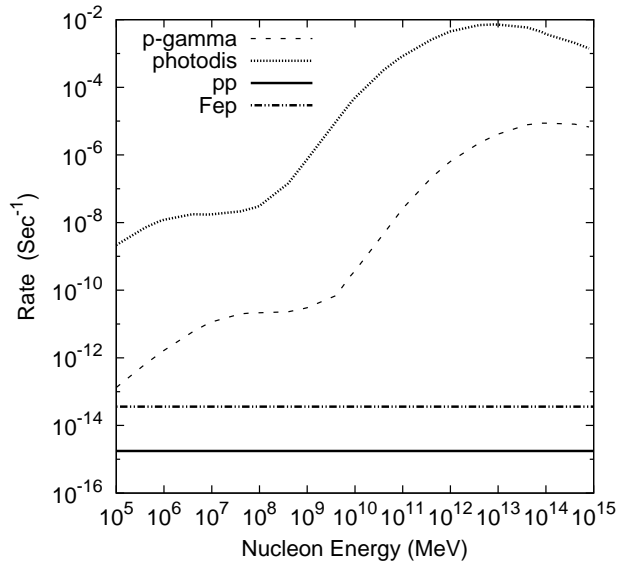}}
\caption{ $p-p$ (solid line), $Fe-p$ (dash-dotted line) for $n_H=1.7cm^{-3}$ \cite{kachel}, $p-\gamma$ (dashed line), photo-disintegration rates of Fe nuclei (dotted line) calculated with the fit of SED \cite{abdo} also given in our FIG.2.
X-axis represents energy per nucleon in the wind rest frame.}
\label{rates}
\end{figure*}
The rate of photo-disintegration of Fe nuclei is the highest among all processes of high energy gamma ray production. The increase in the rates of photo-disintegration and $p\gamma$ interactions near $10^{19} eV$ shown in FIG.1. is due to the first peak or the synchrotron peak in the photon SED as shown in our FIG.2. from \cite{abdo}. The high energy gamma ray flux from photo-disintegration of Fe nuclei is shown with a black dashed line in FIG.2. 
Photo-disintegration of Fe nuclei followed by de-excitation of daughter nuclei
is found to be consistent with the UHECR proton/neutron event rate observed by Pierre Auger between 55EeV, 150EeV and the high energy gamma ray flux measured
by HESS.
\begin{figure*}[t]
\centerline{\includegraphics[width=4.2in, angle=-90]{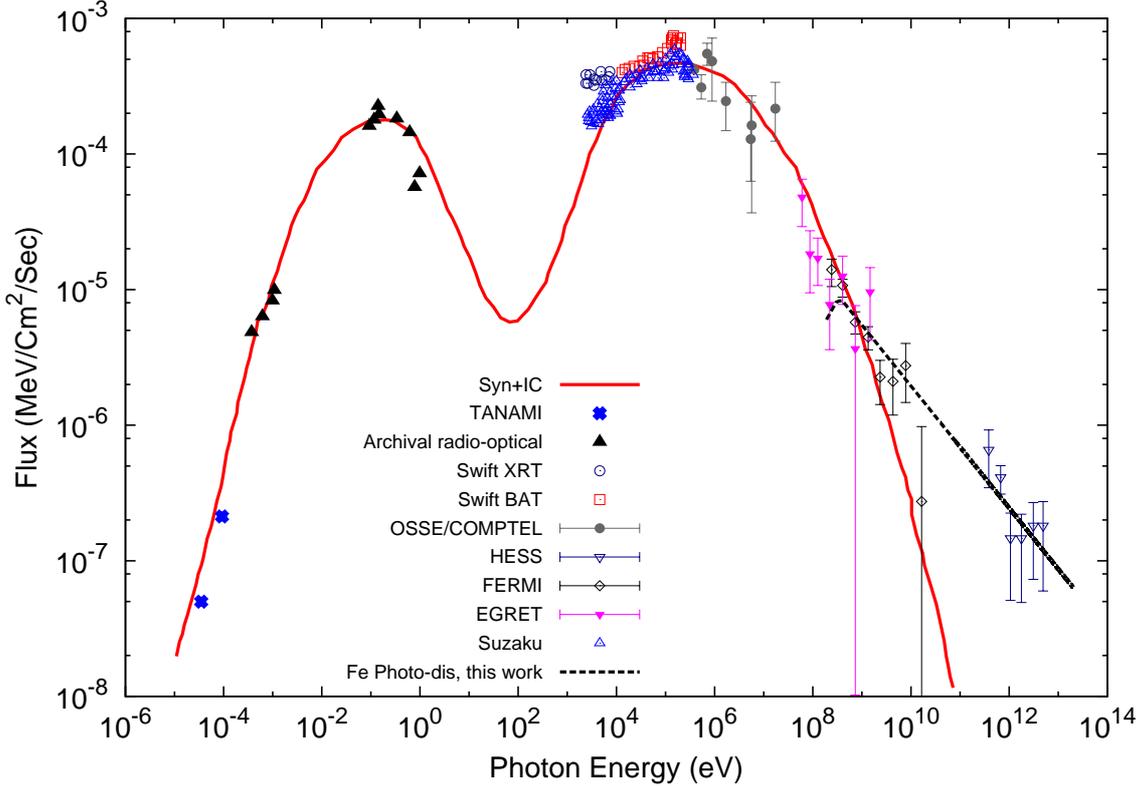}}
\caption{ Spectral energy distribution (SED) ${\epsilon_{\gamma}^o}^2 \frac{dN_{\gamma}^o(\epsilon_{\gamma}^o)}{d\epsilon_{\gamma}^o}(MeV cm^{-2} sec^{-1})$ from Cen A core, the solid red curve is the fit with synchrotron and SSC from \cite{abdo}, high energy gamma ray spectrum from photo-disintegration of Fe nuclei shown with black dashed line.}
\label{SED}
\end{figure*}
 We compare our results with the high energy gamma ray flux estimated in other papers. The $p\gamma$ interaction scenario discussed in the paper \cite{sahu} predicts a peak in the high energy gamma ray spectrum at $190 GeV$ due to the $p\gamma$ interactions of the $170 KeV$ photons with $13 TeV$ protons. Due to photo-disintegration of 
Fe nuclei by 170 KeV photons we expect a peak in the high energy gamma ray spectrum at $400 MeV$.  
 In \cite{fraija} the authors have used the low energy photon spectrum from \cite{abdo} for calculating the high energy gamma ray flux from the lobes of Cen A, but the low energy photon spectrum given in \cite{abdo} is observed from the core of Cen A.     
\par
In summary, we have found that the scenario of $p-p$ intearctions gives excess UHECR events from the core region of Cen A in the energy bin of 55EeV and 150EeV. If we consider there are only Fe nuclei as primary cosmic rays then in the case of pure hadronic interactions $Fe-p$ the estimated UHECR event rate is very low.  Sahu et al. \cite{sahu} have considered the production of 190 GeV gamma rays in interaction of 13 TeV protons with the 170 KeV photons in the second peak of the SED. In their model the luminosity of the 13 TeV protons has to be $4\times10^{45}erg/sec$. In our case $29$ TeV per nucleon energy of Fe nuclei is required to produce gamma rays of energy 250 GeV in photo-disintegration of Fe nuclei. In our model of photo-disintegration of Fe nuclei the total cosmic ray power has to be of the order of $10^{47}$ erg/sec. The required luminosity of the Fe cosmic ray nuclei is higher than the Eddington's luminosity of Cen A. However, we note that the requirement of luminosity depends on the photon density inside the source and size of the emitting region.
 The cosmic ray luminoity required in the photo-disintegration model would be lower if the density of the low energy photons is higher at the source or the size of the emitting region is smaller. 
 Moreover, it has been discussed earlier that the isotropic luminosity in Cen A can easily exceed its Eddington's luminosity which is $10^{46}erg/sec$ during flaring states \cite{dermer}.

\section{Acknowledgment} 
We thank Sarira Sahu and Bing Zhang for helpful communications.

\section{Appendix}
The spectral energy distribution from \cite{abdo} shown by red solid curve in
 our FIG.2. has been fitted in fourteen energy intervals with average error less than $10\%$. The parametrizations used in our calculations are given below.  
\begin{eqnarray}
f(x)=-6.15\times10^{-9}+2.21\times10^3x+2.01\times10^{13}x^2;1.00\times10^{-5}
\leq
\frac{x}{eV} \leq 7.7\times10^{-5}\hskip 2 cm\\
f(x)=1\times10^{-6}-2.06\times10^4x+1.502\times10^{14}x^2;7.7\times10^{-5}
\leq
\frac{x}{eV} \leq 1.17\times10^{-4}\hskip 2 cm\\
f(x)=-1.49\times10^{-7}+5.23\times10^3x+1.49\times10^{13}x^2;1.17\times10^{-4}
\leq
\frac{x}{eV} \leq 4.32\times10^{-4}\hskip 2 cm\\
f(x)=-1.55\times10^{-6}+1.34\times10^4x-4.63\times10^{11}x^2;4.32\times10^{-4}
\leq
\frac{x}{eV} \leq 1.36\times10^{-2} \hskip 2 cm\\
f(x)=5.17\times10^{-5}+3.77\times10^3x-3.99\times10^{10}x^2+1.39\times10^{17}x^3;1.36\times10^{-2}\leq
\frac{x}{eV} \leq 1.34\times10^{-1} \hskip 2 cm\\
f(x)=1.96\times10^{-4}-1.07\times10^2x+2.63\times10^{7}x^2-2.33\times10^{12}x^3;1.34\times10^{-1}\leq
\frac{x}{eV} \leq 4.54 \hskip 2 cm\\
f(x)=5.30\times10^{-5}-5.77x+2.65\times10^{5}x^2-4.14\times10^{9}x^3;4.54\leq
\frac{x}{eV} \leq 28.3 \hskip 2 cm\\
f(x)=3.57\times10^{-6}+2.21\times 10^{-2}x+2.18x^2;2.83\times10^{-2}\leq
\frac{x}{keV} \leq 3.48 \hskip 2 cm\\
f(x)=1.99\times10^{-5}+2.62\times 10^{-2}x-5.74\times10^{-1}x^2;3.48\leq
\frac{x}{keV} \leq 17.8 \hskip 2 cm\\
f(x)=2.14\times10^{-4}+6.44\times
10^{-3}x-5.75\times10^{-2}x^2+1.64\times10^{-1}x^3;17.8\leq \frac{x}{keV} 
\leq 185 \hskip 1 cm\\
f(x)=4.77\times10^{-4}-6.54\times 10^{-5}x+4.21\times10^{-6}x^2;0.185\leq
\frac{x}{MeV} \leq 7.16 \hskip 2 cm\\
f(x)=3.33\times10^{-4}-1.95\times
10^{-5}x+5.55\times10^{-7}x^2-5.31\times10^{-9}x^3;7.16\leq \frac{x}{MeV} \leq 49 \hskip 2 cm\\
f(x)=1.26\times10^{-4}-1.16\times
10^{-6}x+4.35\times10^{-9}x^2-5.49\times10^{-12}x^3;49\leq \frac{x}{MeV} \leq 352 \hskip 2 cm\\
f(x)=2.54\times10^{-5}-3.3\times 10^{-8}x+1.29\times10^{-11}x^2;0.352\leq
\frac{x}{GeV} \leq 1.44 \hskip 2 cm\\
f(x)=2.73\times10^{-6}-2.39\times
10^{-10}x+5.25\times10^{-15}x^2-3.26\times10^{-20}x^3;1.44\leq \frac{x}{GeV} 
\leq
90.94 \hskip 2cm
\end{eqnarray}
\end{document}